\documentclass[a4paper]{iopart}
\usepackage{hyperref}

\begin{document}

\title{How photon astronomy affects searches for continuous gravitational
waves}
\author{Benjamin J Owen}
\address{
Institute for Gravitation and the Cosmos, Department of Physics, The
Pennsylvania State University, University Park, PA 16802-6300, USA
}
\date{$Id: owen-gwdaw13.tex,v 1.16 2009/10/07 17:03:16 owen Exp $}

\begin{abstract}
Due to their computational limitations, searches for continuous
gravitational waves (GWs) are significantly more sensitive when informed by
observational photon astronomy and theoretical astrophysics.
Indirect upper limits on GW emission inferred from photon astronomy
indicate which objects are more interesting for GW searches, and also set
sensitivity milestones which GW searches need to beat to be considered GW
astronomy.
How GW results are interpreted depends on previous indirect limits and the
theory of astrophysical GW emission mechanisms.
I describe the interplay between these issues for the four types of
continuous GW search, and show how photon astronomers can help the growing
field of GW astronomy now and in the near future.
\end{abstract}


\section{Introduction}

The theme of the last two Gravitational Wave Data Analysis Workshops has
been connecting gravitational waves (GWs) to existing astronomy and
astrophysics.
The most commonly discussed connection is between ground-based GW detectors
and observations of electromagnetic counterparts of short-lived GW signals
(binary inspirals or less-modeled bursts)~\cite{Abbott:2008mr}.
But searches for continuous GWs (with ground-based detectors this means
rapidly rotating neutron stars) are even more closely connected to photon
astronomy:
Most searches for continuous waves are computationally limited, and thus
their sensitivity substantially benefits from information on where to look,
even more than inspiral and burst signals.
As a result, we who search for continuous GWs divide neutron stars into four
types, determined mainly by what is already known about them.
For each type of neutron star, photon astronomy also sets indirect upper limits
on GW emission---milestones which GW searches must beat to begin doing GW
astronomy.
Theory has a role too:
Where we look depends on our understanding of GW emission mechanisms, and
our interpretation of our observational results depends on emission
mechanisms and existing indirect limits.
While continuous GW searches present more expensive data analysis problems
than other GW searches, the fact that makes them expensive---the billions
of GW cycles per year---means that they carry very precise information on
their sources and can reveal much about neutron star astrophysics and the
physics of matter at supernuclear densities~\cite{Owen:2009fg,
Sathyaprakash:2009xs}.

Here I review present and near-future LIGO searches for continuous GWs and
place them in their astronomical and astrophysical context.
I skip over most details of the data analysis, which are described in the
observational papers~\cite{Abbott:2003yq, Abbott:2004ig, Abbott:2005pu,
Abbott:2007ce, Abbott:2006vg, Abbott:2007tw, Abbott:2007tda, Abbott:2008fx,
Abbott:2008uq, Abbott:2008rg} and references therein.
I describe the four types of continuous GW searches, sensitivities,
emission mechanisms, and the indirect limits that set sensitivity
milestones.
I focus on issues related to detecting signals rather than extracting
information from them; the latter are summarized
elsewhere~\cite{Owen:2009fg, Sathyaprakash:2009xs}.

\section{Four types of searches}

For the purposes of continuous GW searches, there are four types of neutron
stars: known pulsars (such as the Crab), previously unknown neutron stars,
non-pulsing non-accreting neutron stars (such as the central compact object
in Cas~A), and accreting neutron stars (such as in Sco~X-1).
Therefore I divide continuous GW searches into four corresponding types:
targeted searches with a known timing solution, all-sky surveys, directed
searches for non-accreting stars, and directed searches for accreting stars
(where the orbit and accretion raise extra analysis issues).
This division is made on the basis of differing GW search sensitivities
(strongly affected by what is already known from photon astronomy),
different indirect upper limits on GW emission, and different emission
mechanisms.
In this section I describe the first and last factors, and I describe the
second in later sections devoted to each type of search.

\subsection{Sensitivity}

It is easier to find something if you know where to look, and continuous GWs
are no exception.
This can be seen by comparing observational upper limits on GW strain for
different types of searches---a recent all-sky search~\cite{Abbott:2007tda}
was almost an order of magnitude less sensitive in amplitude than a known
pulsar search of comparable LIGO data~\cite{Abbott:2007ce}, or two orders
of magnitude in GW luminosity (see below).

There are two contributions to the difference:
Fully coherent methods (such as matched filtering) accumulate amplitude
signal-to-noise ratio as $T^{1/2}$, where $T$ is the live time of data
integrated.
Thus a 95\% confidence upper limit on strain tensor amplitude $h_0$ takes
the form
\begin{equation}
\label{h095}
h_0^{95\%} = \Theta \sqrt{S_h/T},
\end{equation}
where $S_h$ is the strain power spectral noise density of the detector at
the GW frequency.
(The generalization to multiple detectors is straightforward but lengthy;
see~\cite{Cutler:2005hc} for the matched filtering version.)

The threshold factor $\Theta$ is determined by the statistics (effective
number of independent trials in parameter space) and efficiency of the
search (coverage of the parameter space) and is the first major
contribution to the difference.
For known pulsars, i.e.\ with coherent timing solutions and precise sky
locations obtained from photon astronomy, $\Theta$ is about
11~\cite{Abbott:2003yq} averaged over sky locations and inclination and
polarization angles.
For the Cas~A search underway, which is directed at a single sky location
but must cover many possible timings, $\Theta$ should be in the
mid-30s~\cite{Wette:2008hg}.

The second contribution to the difference arises because computational cost
can scale very steeply with $T$, especially when searching many sky
locations as in all-sky surveys; and thus the coherent integration time in
the denominator of~(\ref{h095}) is limited.
Semi-coherent methods combine $N$ short (live time $T_\mathrm{coh}$)
coherently integrated stretches of data incoherently, i.e.\ by adding power
not amplitude.
This makes the computation much cheaper, but improves signal-to-noise only
as $N^{1/4} = (T/T_\mathrm{coh})^{1/4}$ (for $N$ more than a few).
Thus for a single detector the sensitivity takes the form
\begin{equation}
h_0^{95\%} = \Theta \sqrt{S_h} (T_\mathrm{coh}T)^{-1/4} = \Theta \sqrt{S_h}
N^{-1/4} T_\mathrm{coh}^{-1/2},
\end{equation}
where for all-sky all-inclination searches the threshold factor $\Theta$ is
in the mid-20s~\cite{Abbott:2007tda}.
(Appendix B of that reference shows how semi-coherent $\Theta$ factors are
typically lower than for comparable coherent searches, e.g.\ 8 instead of 11 for
known pulsars.)

\subsection{Emission mechanisms}

I reviewed continuous GW emission mechanisms fairly
recently~\cite{Owen:2006tz} and space in this article is limited.
Thus here I briefly summarize the main issues and mention only some recent
results which are most relevant to present continuous GW searches and
future searches of accreting neutron stars.
(Specifically I neglect free precession and deformations due to internal
toroidal magnetic fields, as well as most of the $r$-mode story.)

The most interesting mechanism at the moment is rotation of a static
quadrupolar deformation, which can be thought of as a mountain (although
the most important part is generally buried).
The important number is the equatorial ellipticity
\begin{equation}
\epsilon = (I_{xx} - I_{yy})/I_{zz},
\end{equation}
where $I_{ab}$ is the moment of inertia tensor and $z$ is the rotation
axis.
This dimensionless $\epsilon$ is roughly the $\ell=m=2$ mass quadrupole
moment normalized to the moment of inertia, or (if the star had uniform
density) the height of the quadrupolar deformation in terms of stellar radius.
(Radiation from $\ell>2$ and $\ell=2, \,m=1$ is somewhat suppressed.)
Then the GW strain tensor amplitude is
\begin{equation}
\label{quad}
h_0 = (2\pi f)^2 I_{zz} \epsilon / D,
\end{equation}
where $D$ is the distance to the star, the GW frequency is $f=2/P$ in terms
of the spin period $P$, and I have used geometrized units.

The maximum theoretically predicted ellipticity depends on the composition of
the star:
For a normal neutron star (thin solid crust on a liquid interior) the
highest possible is of order $10^{-6}$~\cite{Haskell:2006sv}.
For hybrid stars with partly solid cores due to a baryon-quark or
baryon-meson phase transition spread over a range of densities, the highest
is of order $10^{-5}$~\cite{Owen:2005fn}.
For quark stars, which in some cases are predicted to be mostly solid, the
ellipticity could be $10^{-3}$ or more~\cite{Owen:2005fn, Lin:2007rz,
Haskell:2007zz, Knippel:2009st}.
The maximum ellipticity can be written schematically (neglecting the
integral) as
\begin{equation}
\label{emax}
\epsilon_{\max} = \mbox{(breaking strain)} \times \mbox{(shear modulus)}
\times \mbox{(geometry)}.
\end{equation}
The last factor, which depends on the bulk equation of state and treatment
of hydrostatic equilibrium, is uncertain by a factor of a few (as seen by
properly comparing~\cite{Haskell:2006sv} to older results).
The shear modulus is the main reason for the orders of magnitude variation
between models, as it increases rapidly with density and thus is highest in
exotic phases which are solid at the highest densities,
e.g.~\cite{Mannarelli:2007bs}.
The breaking strain has also been considered uncertain by orders of
magnitude, with most $\epsilon_{\max}$ results quoting a highest value of
$10^{-2}$ (roughly the maximum seen on Earth).
However recent simulations~\cite{Horowitz:2009ya} indicate that neutron
star crust, due to the great pressure, is a nearly perfect crystal with
breaking strain of order $10^{-1}$.
Then the maximum ellipticity of a normal neutron star is ten times higher
($10^{-5}$), and if the effect holds for exotic cores the other maximum
ellipticities are also ten times higher.

A separate issue is how to drive $\epsilon$ toward $\epsilon_{\max}$.
In young neutron stars it is not clear what would cause large sustained
elastic deformations or how long they would last, although the violence and
asymmetry of the supernova explosion may help with the former.
But for accreting neutron stars there are several mechanisms to drive and
sustain asymmetry (which must exist as shown by x-ray pulses and burst
oscillations).
Lateral temperature gradients due to non-spherical accretion flow affect
electron capture rates and thereby effectively lift denser material in
hotter regions (see~\cite{Watts:2008qw} and references therein).
On the surface, the magnetic field will funnel incoming accreted material
onto the magnetic poles, e.g.~\cite{Vigelius:2009yn} and references therein.
Accretion can also, through a complicated set of physical effects briefly
surveyed in parts of~\cite{Sathyaprakash:2009xs}, sustain the $r$-modes
(Coriolis-dominated fluid oscillations) of a neutron star.
In this case the continuous GW frequency is $fP \approx 4/3$ rather than
$fP=2$, and $h_0$ is a function of mode amplitude rather than static
ellipticity.
($R$-modes may also be active for a time in young neutron stars.)

\section{Known pulsars}

With a precise timing solution and sky location, a pulsar's spin frequency,
frequency evolution (including glitches and timing noise), and Doppler
shifts due to the detectors' motion and pulsar's binary motion (if any) are
all known and can be de-modulated.
Then it is computationally feasible to coherently integrate the entire data
set---S5, the longest so far~\cite{Abbott:2007kva} is about two calendar
years of LIGO data (one year triple coincidence between three
detectors)---to achieve the optimal signal-to-noise ratio and make the most
of this hard-won data.

With $fP=2$, the current LIGO low-frequency cutoff of 40~Hz means that
pulsars with $P<50$~ms are of interest; and with advanced LIGO (and Virgo)
this will extend to $P<200$~ms~\cite{advanced, Acernese:2008zzf}.
(All searches so far have assumed $fP\approx2$, but future searches will
also investigate $fP\approx4/3$ and $fP\approx1$.)
Thus we are looking for young fast pulsars and recycled millisecond
pulsars, but not garden-variety pulsars, rotating radio transients,
anomalous x-ray pulsars or soft gamma repeaters.
At the moment there are almost 200 known pulsars in the LIGO band, or about
10\% of the total~\cite{2005AJ....129.1993M}.

For known pulsars the common indirect limit on GW emission is based on the
observed spin-down.
Given the observed period and spin-down (period derivative), the assumption
that all the spin-down is due to GW emission sets a robust upper limit on
$h_0$.
That limit is found (for a static $m=2$ quadrupole) to be
\begin{equation}
h_0^\mathrm{sd} = 4.5\times10^{-24} \left( 1\mathrm{~kpc} \over D \right)
\left( I_{zz} \over 10^{45}\mathrm{~g\,cm}^2 \right)^{1/2} \left(
10^3\mathrm{~yr} \over P/\dot{P} \right)^{1/2}.
\end{equation}
Due to uncertainties in the moment of inertia and the distance, this limit
is uncertain typically by a factor of 2.
The highest spin-down limit in the current LIGO band is for the Crab
pulsar, with fiducial value $h_0^\mathrm{sd} = 1.4\times10^{-24}$.
The Vela pulsar has a comparable $h_0^\mathrm{sd}$ and will be in the band
of advanced LIGO and Virgo.
A search for Vela in Virgo data, although not yet at full sensitivity, is
underway.

Of course the true GW emission is less than the spin-down limit; and if we
know other things about the pulsar we can set limits on how much less.
The emission of the pulsar itself is typically less than 1\% of the
spin-down luminosity, but for young pulsars associated with a pulsar wind
nebula we can subtract off the bolometric luminosity of the nebula (which
may be tens of percent).
In some cases the second period derivative or equivalently braking index is
known; and then a stricter indirect limit can be computed using a
phenomenological model of the star's spin-down
history~\cite{Palomba:1999su}.
For the pulsars examined, this limit on $h_0$ is a few times stricter than
the spin-down limit.

The spin-down limit on $h_0$ corresponds to an ellipticity upper limit
\begin{equation}
\epsilon_\mathrm{sd} = 1.1\times10^{-4} \left( I_{zz} \over 10^{45}
\mbox{~g\,cm}^2 \right) \left( 10^3\mathrm{~yr} \over P/\dot{P}
\right)^{1/2} \left( P \over 10\mathrm{~ms} \right)^2 .
\end{equation}
For young pulsars this limit is typically of order $10^{-4}$ or higher (for
the Crab it is $7\times10^{-4}$), while for recycled millisecond pulsars it
can be $10^{-10}$ or lower.

The first LIGO continuous-wave search~\cite{Abbott:2003yq} of the first
science data run (S1) focused on a single pulsar, J1939+2134.
(At the time it was the most rapidly rotating known pulsar, thus maximizing
$h_0$ for a given ellipticity.)
The next search was expanded to 28 pulsars in S2 data~\cite{Abbott:2004ig}.
The author list includes not only the LIGO Scientific Collaboration (LSC)
but Michael Kramer and Andrew Lyne, radio astronomers who obtained timing
data crucial to the sensitivity of the search.
With the combined S3 and S4 data set the total rose to 78
pulsars~\cite{Abbott:2007ce}, and again Kramer and Lyne were co-authors for
obtaining crucial pulsar timing.
The best upper limit on $h_0$ from the S3/S4 search was $3\times10^{-25}$
for a pulsar near the minimum of the noise spectrum using several weeks
each of S3 and S4 data.
All of these searches employ fully coherent methods, since for known
pulsars they are computationally feasible.

Most recently the LSC published a search~\cite{Abbott:2008fx} of the first
9 months of S5 for the Crab pulsar which beat the spin-down limit and the
stronger indirect limits from the braking index and luminosity of the
pulsar wind nebula.
The best upper limit on GW emission---which also takes advantage of the
inclination of the pulsar spin axis inferred from x-ray images of the
nebula---was $h_0 = 3\times10^{-25}$ or about 4\% of the canonical
spin-down luminosity, beating the tightest indirect limit on $h_0$ (from
the braking index) by a factor of 2.
Also included was a broader search allowing for a possible difference
between the radio timing and GW signal behavior, motivated by the
possibility that GW and photon emission might come from separate components
of the star and by the fact that glitches indicate the existence of
multiple components rotating at (sometimes) different rates.
The energy upper limit of this latter search was more than an order of
magnitude worse than the former search due to the effects of searching over
many possible timings.

The best ellipticity upper limit on the Crab was $1\times10^{-4}$, within
the range possible for stars with crystalline quark cores.
It is very interesting that this is below $\epsilon_{\max}$ for various
exotic matter models.
However, I must correct a misconception~\cite{Lin:2007rz, Haskell:2007zz}
which is spreading~\cite{Knippel:2009st} that LIGO upper limits are
constraining QCD parameters:
A GW upper limit on $\epsilon$ does not constrain anything unless it beats
$\epsilon_\mathrm{sd}$ (not done until the Crab
search~\cite{Abbott:2008fx}), and even then it only constrains $\epsilon$
of that star (a function of its history) rather than $\epsilon_{\max}$ (a
function of universal properties of dense matter):
$\epsilon$ may be much less than $\epsilon_{\max}$ simply because the
actual strain (as opposed to the breaking strain) is very small.
Constraints will only come with many more observations beating spin-down
limits as well as work on mountain building theory~\cite{Owen:2005fn}.

At the time of writing, a search of all available S5 data for the Crab and
more than 100 other pulsars is nearing completion, and will be published by
the LSC and Virgo Collaboration with several more radio and x-ray
astronomer co-authors.
Coherent timing solutions provided by the latter were crucial for making
the most of the GW data.
Not all of the pulsars in the LIGO/Virgo band are included due to a lack of
coherent timing solutions across the S5 run for some pulsars.

The GW collaborations are eager to obtain timing for every pulsar in band,
especially the ones with spin-down limits near what is achievable.
Values of $h_0^\mathrm{sd}$ can be inferred from the ATNF
catalogue~\cite{2005AJ....129.1993M} and compared to the generic pulsar
sensitivity estimate~(\ref{h095}) assuming one year of triple coincident
data at present or advanced noise levels.
Due to varying angles and the intrinsic randomness of the noise, this
sensitivity estimate is uncertain at about the factor of 2 level
characteristic of the spin-down limits.
Therefore even pulsars with fiducial spin-down limits within about a factor
3 of~(\ref{h095}) are extremely interesting.
(The others are still interesting in case of extreme luck, such as the
dispersion-measure distance being badly overestimated.)
With present data the most interesting pulsars after the Crab are
J0537-6910, J1952+3252, and J1913+1011.
Later this year LIGO is scheduled to start another long science run
(``enhanced LIGO'') with strain noise amplitude reduced by about
2~\cite{enhanced}, and Virgo is scheduled for an enhanced run on a
similar timescale.
In the era of advanced LIGO (strain noise an order of magnitude lower than
S5 data) more than 70 fiducial spin-down limits will be beatable
by~(\ref{h095}), and of order 100 will be close to it.

\section{Unknown neutron stars}

Most of the neutron stars in the galaxy are not seen via any form of
photons, but they may still emit continuous GWs.
We must then search over the whole sky, the frequency band of the
detectors, and possible frequency evolutions of the sources; and even using
every computer on the planet still could not coherently integrate a year's
worth of data in a year.
Thus we need to use semi-coherent data analysis methods which trade off
computational cost for sensitivity.

Even for the many unseen neutron stars in the galaxy we can place an indirect
limit on $h_0$.
The original version of this limit was due to Blandford in the 1980s
(unpublished), but has reappeared several times in the literature, most
recently in~\cite{Abbott:2006vg}.
With a galactic neutron star birth rate $R$ derived from supernova
observations and an infinite planar spatial distribution, the
supernova-based indirect limit (which is a statistical statement about the
brighest source) is
\begin{equation}
h_0^\mathrm{SN} = 4\times10^{-24} \left( R \times 30\mathrm{~yr} \right),
\end{equation}
completely independent of the typical ellipticity.
A more recent treatment~\cite{Knispel:2008ue} based on realistic spatial
distributions of neutron stars produces a more complicated result, which
does depend on the typical ellipticity.
The highest values of $h_0^\mathrm{SN}$ are about $1\times10^{-24}$ for an
extremely optimistic but not \textit{a priori} impossible set of
assumptions.

The most basic all-sky continuous GW survey was a fully coherent matched
filtering search of 10 hours of LIGO S2 data~\cite{Abbott:2006vg}.
Semi-coherent searches were published for S2~\cite{Abbott:2005pu} and
recently for S4~\cite{Abbott:2007tda}, both using several weeks of data
with a coherent integration time of half an hour.
The best strain upper limit from the latter was $4\times10^{-24}$
(averaging over inclinations as in the known pulsar limits from comparable
data), more than an order of magnitude in amplitude (i.e.\ two orders in
luminosity) worse than the best S3/S4 known pulsar limit of
$3\times10^{-25}$~\cite{Abbott:2007ce}.
Also, the all-sky searches used much more computing power than the known
pulsar search, again illustrating the advantages of knowing where to look.
The S4 all-sky searches (three were reported in~\cite{Abbott:2007tda})
nonetheless achieved a sensitivity comparable to Blandford's indirect
limit.
The S4 searches also achieved non-trivial range milestones, as described
and plotted toward the end of that paper.

More recently, the most sensitive all-sky search to date was published on
several months of the early S5 data~\cite{Abbott:2008rg}, again with a
half-hour coherent integration time.
This search placed upper limits on GW emission ($h_0$ as low as about
$1\times10^{-24}$ for an average inclination) ruling out the most extreme
versions of the improved supernova indirect limit~\cite{Knispel:2008ue}.
Another version of this search on a greater fraction of S5 data is
underway, and is expected to yield even greater sensitvity.

The ultimate solution to the computational limitation of the all-sky
searches is Einstein@Home~\cite{EAH}, a distributed computing project
involving tens of thousands of users and otherwise idle computers from
around the world.
This enormous computing power allows the coherent integration time to be
increased to more than a day, which is still small compared to the months
of coherent integration for known pulsars.
Recently a search of S4 data was published~\cite{Abbott:2008uq}
demonstrating the power of Einstein@Home, although that search was not
designed to place upper limits.
Since that search, Einstein@Home has been running improved searches on S5
data, the first of which was recently published~\cite{Abbott:2009nc}.

All these all-sky surveys have searched for isolated neutron stars.
However plans are being made to search for unseen neutron stars in
binaries, such as faint quiescent low-mass x-ray binaries, after data
analysis algorithms are further developed to handle the extra challenge of
searching over unknown binary parameters.

\section{Non-pulsing non-accreting neutron stars}

For some non-accreting neutron stars the position is known, but the spin
frequency and its evolution are unknown.
The prototype is the central compact object in supernova remnant Cas~A.
With some knowledge of where to look in parameter space, the sensitivity of
such searches should be intermediate between those for known pulsars and
unseen neutron stars.

For these neutron stars we do not know the spin-down and thus cannot
use the spin-down limit.
However if we make the same physical assumption, that the spin-down is
dominated by quadrupolar GW emission, and additionally assume that this has
always been so (and that $I_{zz}$ has remained constant), we obtain an
indirect strain limit based on the age $a$,
\begin{equation}
h_0^\mathrm{age} = 2.3\times10^{-24} \left( 1\mathrm{~kpc} \over D \right)
\left( I_{zz} \over 10^{45} \mathrm{~g\,cm}^2 \right)^{1/2} \left(
10^3\mathrm{~yr} \over a \right)^{1/2}.
\end{equation}
The highest such limit which is well determined is for Cas~A at
$1.2\times10^{-24}$~\cite{Wette:2008hg}.
The corresponding ellipticity limit is $4\times10^{-4}$--$4\times10^{-5}$
over the frequency band 100--300~Hz.
There is one big caveat not present with known pulsar searches:
We have to hope that the GW emission is in the LIGO/Virgo band, which is
true only for about 10\% of all known pulsars, and a smaller fraction of
young isolated pulsars.

The Cas~A search underway~\cite{Wette:2008hg}, the first of this kind, is
fully coherent over 12 days of LIGO S5 data and covers the frequency band
100--300~Hz and many values of $\dot{f}$ and $\ddot{f}$.
It should beat $h_0^\mathrm{age}$ over that band, with the best
$h_0^\mathrm{95\%}$ about $8\times10^{-25}$ at the minimum of $S_h$, and
cost more than the known pulsar searches but substantially less than
all-sky surveys.
This will add to the very short list of objects where a GW search
(continuous-wave or otherwise) has beaten an indirect limit from photon
astronomy.

There are many interesting possibilities beyond Cas~A~\cite{where}.
Several other supernova remnants are inhabited by similar non-pulsing
objects.
``Empty'' supernova remnants and pulsar wind nebulae are also attractive
places to do this type of GW search, as long as they are not too large or
too old.
(With too many sky positions, the sensitivity of the search will not beat
the all-sky surveys, and with older objects the neutron star may have been
kicked far enough to be hard to find.)
Massive star forming regions are also good targets for hunting young
neutron stars, and globular clusters with dense cores are attractive places
to look for older neutron stars which have been effectively rejuvenated by
close encounters.
Directed searches are being planned for all these targets, and work is also
in progress on more sensitive data analysis techniques
(e.g.~\cite{resampling}) which will prove especially fruitful for this type
of search.

\section{Accreting neutron stars}

Accreting neutron stars may be the best continuous GW sources because they
are driven toward the maximum asymmetry, whether ellipticity or $r$-mode.
These stars are most likely to emit continuous GWs at the indirect limit,
but that limit is lower than for other types:
For accreting neutron stars the indirect limit is derived by assuming that
accretion spin-up and GW spin-down are in equilibrium, and thus the most
rapidly accreting stars (i.e.\ those in low-mass x-ray binaries) are the
most interesting.
Taking the observed x-ray flux $F_x$ as a proxy for the accretion rate, we
have for a neutron star of mass $M$ and radius $R$
\begin{eqnarray}
h_0^\mathrm{acc} &=& 3\times10^{-27} \left( F_x \over
10^{-8}~\mathrm{erg\,cm}^{-2}\,\mathrm{s}^{-1} \right)^{1/2} \left( R \over
10~\mathrm{km} \right)^{3/4}
\nonumber \\ &&\times
\left( 1.4~M_\odot \over M \right)^{-1/4} \left( P\times 300~\mathrm{Hz}
\right)^{1/2}
\end{eqnarray}
assuming emission from ellipticity~\cite{Watts:2008qw}.
(For an $r$-mode $h_0^\mathrm{acc}$ is about 1.5 times greater at fixed
$P$.)
The highest such limit is for Sco~X-1 at $3\times10^{-26}$ (for the
fiducial parameters), which is still two orders of magnitude lower than the
largest indirect limits on continuous GW emission for the other types of
neutron star.

The LSC has published two searches for Sco~X-1 so far.
The fully coherent search of S2 data~\cite{Abbott:2006vg} was restricted to
only 6 hours of data, while the semi-coherent S4
search~\cite{Abbott:2007tw} used several weeks of data divided into
one-minute coherent intervals.
The latter search obtained a best upper limit on $h_0$ of
$3\times10^{-24}$, more than two orders of magnitude weaker than the
indirect limit.
The coherent integration time is much shorter than for Cas~A primarily due
to uncertainties in orbital parameters, and on longer timescales the
stochasticity of accretion torque would also come into play.

In contrast to the other types of continuous GW search, no searches for
accreting neutron stars are underway.
This is because the indirect limits will be accessible only with advanced
LIGO and Virgo.
The widespread impression to the contrary is due to the fact that the
limits would be accessible with a known timing solution.
In general the highest indirect limits are for the few ``bright Z'' sources
accreting near the Eddington limit.
Unfortunately these are also the sources for which the spin frequency is
completely unknown, complicating the GW data analysis.
And the accreting stars with best timing information have the lowest
average accretion rates and thus indirect GW limits that are too low even
with the advantages of known timing~\cite{Watts:2008qw}.

However in the long run accreting neutron  stars may be the best prospects
for GW detection rather than upper limits, because of numerous mechanisms
through which accretion can drive the asymmetry needed to emit GWs.
Therefore the LSC and Virgo collaboration are designing better data
analysis techniques, for example based on the frequency comb used in radio
pulsar searches~\cite{Messenger:2007zi}.
It is also important to consider astrophysical models beyond the strict
torque-balance scenario:
Torques might be balanced only on average over long timescales, with GW
active and quiescent episodes; and during the former the GW signal could be
stronger than $h_0^\mathrm{acc}$~\cite{Watts:2009gk}.
Upper limits can then constrain such scenarios.

\section{Present and future observational interactions}

I close by summarizing the ways that observational photon astronomy is
already helping continuous GW searches and can do so in the future.

The most sensitive continuous GW searches are for known pulsars.
It is imperative for the ground-based GW detection collaborations to
maintain and expand links to the pulsar timing community, ideally obtaining
timing for all pulsars in the LIGO/Virgo band during the data runs.
Discovery of more pulsars in the band with high spin-downs will also be
extremely helpful, with radio projects culminating in the Square Kilometer
Array expected to expand the list of known pulsars by an order of
magnitude.
Sensitive, frequent, and flexible x-ray timing (such as provided by the
aging Rossi X-ray Timing Explorer satellite) is also important as the only
way to obtain up-to-date parameters on PSR J0537-6910, one of the most
interesting (glitchy, high spin-down limit) pulsars.
As shown by the LIGO Crab result~\cite{Abbott:2008fx}, studies of pulsar
wind nebulae revealing the inclination angle can also improve the
sensitivities of GW searches.

Discoveries of new non-pulsing neutron stars will add interesting targets
to the list of directed searches, which are more sensitive than all-sky
surveys.
It is helpful to find spin periods, which move these objects to the
category of known pulsars---as long as the spin period is short enough for
$2/P$ to be in the LIGO/Virgo band.
Other observations (e.g.\ to better determine ages and distances) of
existing objects will help rank targets in terms of indirect limits on GW
emission.
For this purpose it is also helpful to find pulsar wind nebulae (with or
without a compact object), which are correlated with high spin-downs.

Several years from now, searches for continuous GWs from accreting neutron
stars will reach interesting sensitivities.
They would be aided greatly by determination of spin periods of bright Z
sources, as well as by continued timing of accreting millisecond pulsars
and spin (and orbital) period estimates from burst oscillations.

Last but certainly not least, anyone can help the all-sky surveys by running
the Einstein@Home screensaver~\cite{EAH}, which now searches Arecibo radio data
for new pulsars as well.

\ack
This work was supported by NSF grant PHY-0555628.

\section*{References}
\bibliographystyle{unsrt}
\bibliography{owen-gwdaw13}

\end{document}